\documentclass[twocolumn,showpacs,prb,amsmath,nofootinbib,amssymb]{revtex4-1}
\usepackage{color}
\usepackage{graphicx}
\usepackage{times}

\begin{document}
\title{Hybridization of Graphene and a Ag Monolayer Supported on Re(0001)}

\author{M. Papagno$^{\ast1,2}$, P. Moras$^{1}$, P. M. Sheverdyaeva$^{1}$, J. Doppler$^{3}$, A. Garhofer$^{3}$, F. Mittendorfer$^{3}$, J. Redinger$^{3}$, and C. Carbone$^{1}$}
\affiliation{
$^{1}$Istituto di Struttura della Materia, Consiglio Nazionale delle Ricerche, Trieste, Italy \\ 
$^{2}$Dipartimento di Fisica, Universit$\acute{a}$ della Calabria, 87036 Arcavacata di Rende (Cs), Italy \\ 
$^{3}$Institute of Applied Physics and Center for Computational Material Science, \\
Vienna University of Technology, Gusshausstrasse 25/134, A-1040 Vienna, Austria
}
\begin{abstract}
We have investigated the electronic structure of graphene supported on Re(0001) before and after the intercalation of one monolayer of Ag
by means of angle-resolved photoemission spectroscopy measurements and density functional theory calculations.   
The intercalation of Ag reduces the graphene-Re interaction and modifies the electronic band structure of graphene. 
Although the linear dispersion of the $\pi$ state of graphene in proximity of the Fermi level highlights a rather weak graphene-noble metal layer interaction, 
we still observe a significant 
hybridization between the Ag bands and the $\pi$ state in lower energy regions. These results 
demonstrate that covering a surface with a noble metal layer does decouple the electronic states, but still leads to a
noticeable change in the electronic structure of graphene.
\end{abstract}

\maketitle

\section{Introduction}

A detailed understanding of the chemical interaction between graphene and a metal substrate is a major prerequisite for tailoring the electronic properties of graphene, as it allows tuning
the electronic states of graphene by changing the support or via the intercalation of alloy materials~\cite{bat12}.
\newline The interaction strength of graphene with late transition metals strongly depends on the choice of the support. In the case of  Pt~\cite{zi87, su09}, Ir~\cite{dia06,ple09} and Cu~\cite{gao10, wal11}, 
only a weak interaction is found, where the graphene layer is located at about 3.7~\AA~from the metal surface. 
On these substrates, the interaction only leads to minor changes in the local density of states, but the key electronic properties of graphene such as the linear dispersion 
at the Dirac point are mostly preserved.  
\newline On the other hand, graphene adsorbed on Ni~\cite{gam97}, Rh~\cite{wan10} and Ru~\cite{mor10, str11} is found at a minimum distance of about 2.1~\AA~ and the interaction is stronger. 
For these systems, the hybridization between the carbon and metal atoms leads to a loss of the linear dispersion of the graphene bands~\cite{nag94, bru09, pac12}.  
In these cases the electronic properties can be restored by intercalation of noble metal atoms, as evidenced by several angle-resolved photoemission spectroscopy (ARPES) studies~\cite{end10, var10, pop11}.
The intercalated noble metal layers do not only act as spacers, but also reduce the hybridization between the metal {\it d} orbitals and graphene $\pi$ band.  
Indeed, a stiffening of the graphene phonon modes~\cite{shi99, far99, shi00} and the recovered linear dispersion of the $\pi$ band have been taken as indication
that graphene is only marginally perturbed.
However, the exact role of the intercalated film is not well understood.
In this study, we show that even as the noble metal film leads to a decoupling of the substrate, the electronic states of  noble metal layer still hybridize with the graphene layer, 
and induce a band gap in the graphene $\pi$ band. 
\newline
For a detailed analysis, we investigated the electronic structure of graphene supported on Re(0001) before and after the intercalation of one monolayer (ML) Ag film with ARPES experiments and density functional theory 
(DFT) calculations to assess the decoupling of the graphene sheet.  
Due to the lattice mismatch, graphene on Re(0001) forms an incommensurate phase that exhibits a moir\'e pattern consistent with a (10$\times$10)-graphene unit cell over a (9$\times$9)-Re(0001) unit cell~\cite{min11}. 
Similar to Ru(0001) and Rh(111)~\cite{pre08}, we find a strong graphene-substrate interaction on the bare Re(0001) surface, as indicated by  a splitting of the C 1{\it s} photoemission peak of graphene 
and the electronic structure measured  from ARPES experiments. In the following step, the intercalation of Ag restores the linear dispersion of the $\pi$ band in the vicinity of the Dirac point.
However, the measurements show that the interaction with the Ag induced electronic states at 4-7~eV below  the Fermi level ({\it E$_{F}$}) leads to the formation of a band gap in the graphene $\pi$ band, in agreement
with the predictions from the DFT calculations.

\begin{figure*}
\includegraphics[width = 2\columnwidth]{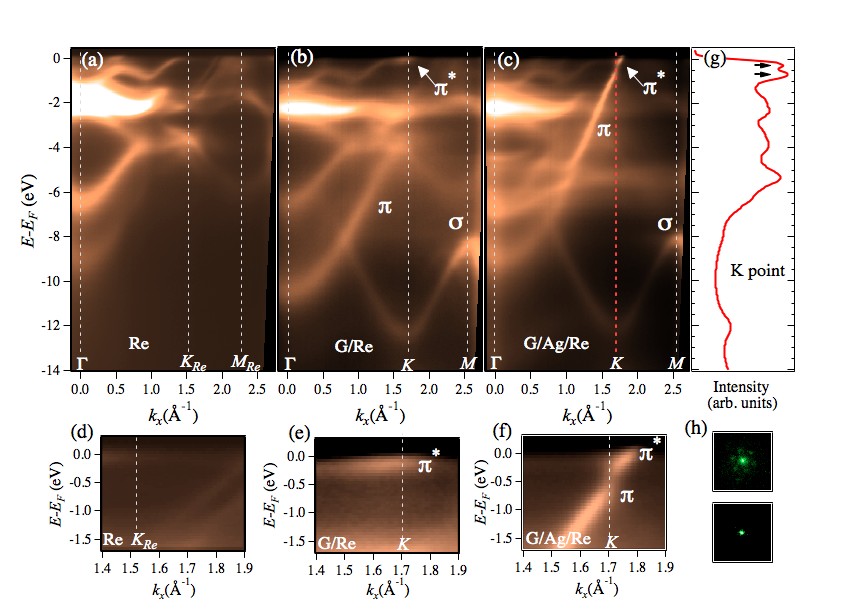}
\caption{(Color online) (a) Photoemission intensity plot along the $\Gamma$${\rm K}$ direction of Re(0001), (b) of graphene/Re (G/Re), and (c) after intercalation of one monolayer of Ag (G/Ag/Re).  Panels (d), (e), and (f) display magnified views of (a), (b) and (c), respectively. (g) Energy distribution curve measured at the ${\rm K}$ point of graphene at G/Ag/Re surface (red dashed curve in panel c) evidences  a band gap highlighted by arrows.  In order to increase the signal/noise ratio, the energy distribution curve has been obtained by summing curves with momenta $\pm$0.004~\AA$^{-1}$~around the ${\rm K}$ point.  (h) LEED pattern measured at 55~eV displaying the (0, 0) spot for G/Re (top) and G/Ag/Re (bottom) surfaces.
}
\label{arpes}
\end{figure*}

\section{Methods}
The photoemission experiments were carried out at the VUV-Photoemission beamline of the Elettra synchrotron radiation facility. ARPES experiments were performed at 120~eV photon energy  at room temperature, using a Scienta R4000 electron energy analyzer with an angular aperture of 30$^{\circ}$ and energy resolution of 20~meV. The electronic band structure and high-symmetry points of each surface have been identified by collecting ARPES measurements at different azimuthal angles over an angular range of more than 40$^{\circ}$. 
The Re(0001) sample was cleaned by repeated cycles of oxidation followed by annealing at 2000~K. Graphene was grown by dosing 30~L (1~L  corresponds to 1.33$\cdot10^{-6}$~mbar for 1~s) ethylene (C$_{2}$H$_{4}$) with the sample temperature  held at 1100~K. 
Intercalation of Ag is achieved by depositing 1 ML of Ag on graphene/Re,  followed by annealing at 500~K. During the annealing some of the Ag atoms 
intercalate, whereas  others, most likely coalesce to form clusters. In this condition, both the parabolic and the linear dispersive $\pi$ states are observed. 
The deposition/annealing process is iterated until no trace of parabolic $\pi$ band is observed. Further Ag deposition and annealing does not lead to a change in the electronic structure suggesting that the intercalation process may be limited to a single-layer of noble atoms, in agreement with earlier results~\cite{var10, end10, pop11, sta04}.

DFT calculations were performed with the Vienna {\it ab-initio} simulation package (VASP)~\cite{kre93,kre96}. 
We employed PAW potentials~\cite{blo94,kre99} and an energy cutoff of 400 eV. The van der Waals density functional
optB88-vdW~\cite{kli11} 
was applied to approximate the exchange-correlation potential, as it allows to capture the non-local contributions
to the adsorption of graphene on a metal surface \cite{mit11}.
The DFT lattice constants of  $a=2.777$~\AA~and $c=4.482$~\AA~were used for the Re substrate.
Due to the slightly different ratio of the theoretical lattice constants, the calculations were carried out for  a ($9 \times 9$)-graphene layer supported  by 3 ($8 \times 8$)-Re layers and an epitaxial Ag layer, 
resulting in a strain of only 0.2 \% in the graphene layer ($a=2.463$~\AA).
In addition, we have investigated  a smaller model consisting of a ($1\times1$) unit cell using a six layer slab with the uppermost three
layers relaxed.
To integrate the Brillouin zone a $\Gamma$-centered ($15\times15\times1$)  {\it k} point mesh was used for the primitive
cell and a ($3\times3\times1$) mesh for the larger cell. In addition, the role of many-body effects arising from the {\it e-e} interactions for the band structure of the intercalated noble metal has been explored 
in the framework of the G$_0$W$_0$ approximation \cite{shi06}. 
The band structure of the large supercell was unfolded by projecting on the first Brillouin zone of primitive (1$\times$1) graphene 
unit cell via an explicit evaluation of  the overlap of the wavefunction at the reciprocal vector $k$ in the primitive cell and a corresponding   
$k^\prime$ in the extended large cell.

\section{Results and Discussion}


Figure~\ref{arpes} shows the experimental electronic band structure along the $\Gamma$${\rm K}$ direction of bare Re (Figure~\ref{arpes}(a)) and graphene/Re (Figure~\ref{arpes}(b)). The formation of graphene is evidenced by the electronic states  $\pi$ and $\sigma$. 
In contrast to the conical shape  of the $\pi$ bands dispersing linearly towards the Fermi level~\cite{ple09,  wal11, rus10, pap12} displayed by weakly interacting graphene,
for graphene/Re(0001) we observe a  parabolic dispersion of the $\pi$ band with a maximum at the  K point 3.90~eV below {\it E$_{F}$}. 
Similar to graphene/Ru(0001)~\cite{bru09, end10, papa12},  the hybridization with the metal {\it d} states modifies the $\pi^{\ast}$ state of graphene into a diffuse band (indicated by the arrow). These observations provide clear evidence for the strong interaction between graphene and the Re substrate.

\begin{figure*} [ht]
\includegraphics[width = 2\columnwidth]{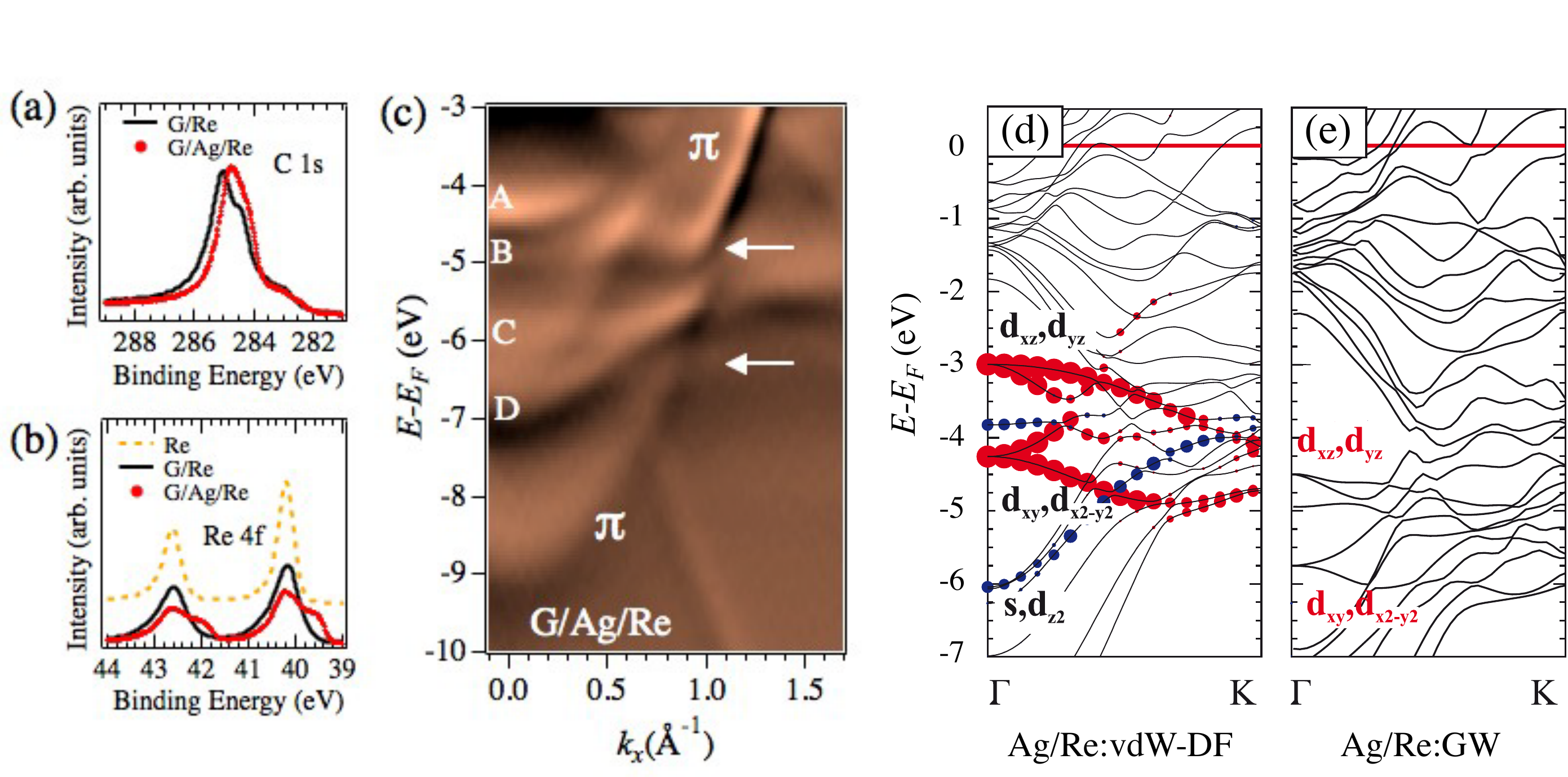}
\caption{(Color online) (a) C 1{\it s} photoemission peak measured at normal emission with photon energy of 410~eV for G/Re (black solid curve) and 
G/Ag/Re (red dotted curve).  (b) Re 4{\it f} states for clean Re (yellow dashed curve), G/Re (black solid curve), and G/Ag/Re (red dotted curve) collected with photon energy of 650~eV at normal emission.  Re 4{\it f} spectrum for clean Re has been vertically shifted for the purpose of comparison. (c) First derivative ARPES map  of G/Ag/Re  along the $\Gamma$${\rm K}$ direction in the energy range 3-10~eV below E$_{F}$. A, B, C and D denote new 
features in the electronic band structure due to Ag intercalation. The arrows highlight the induced electronic gap in the graphene $\pi$ band.
(d) Calculated (vdW-DF) band structure of a monolayer Ag/Re  along the $\Gamma$${\rm K}$ direction (dot size  indicates
the localization in the Ag layer), and (e) $G_0W_0$ band structure along the $\Gamma$${\rm K}$ direction for Ag/Re.
}
\label{arpes&cores}
\end{figure*}

Figure~\ref{arpes}(c) displays the electronic band structure of graphene/Re after the intercalation of Ag. To better visualize the electronic states close to the Fermi level we also show in panels (d), (e) and (f) magnified views of the ARPES maps displayed in (a), (b) and (c) respectively. 
The van der Waals DFT calculations for a ($9 \times 9$) graphene layer supported  on a ($8 \times 8$) Ag/Re
substrate already indicate a rather weak adsorption: with an average adsorption energy of 60~meV/C, 
the graphene sheet is located at an average distance of 3.47~\AA~from the Ag layer and only a minor  buckling of 0.21~\AA~
in the graphene layer is predicted. 
The reduced graphene-substrate interaction is confirmed experimentally by the shift of the $\pi$ state~\cite{sta04, var10, end10}, in our case by about 1.60~eV towards  {\it E$_{F}$} at the $\Gamma$ point and by the linear dispersion in the proximity of the Fermi level. 
The electron charge transfer from Ag to graphene shifts the Dirac point to an energy of 0.4~eV below the Fermi level, allowing for the observation of  the $\pi^{\ast}$ band,
compared to a slightly smaller calculated value of 0.1 eV below the Fermi level.  
A similar shift has been predicted by  DFT for the adsorption of graphene on a {\it bare} Ag surface~\cite{gio08}. 
Although the DFT bandstructure  for the full model does not show  a bandgap at the {\it K} point, the experimentally measured distribution curve displayed in 
Figure~\ref{arpes}(g) reveals an energy gap of (0.45$\pm$0.10)~eV. Comparable gaps have been reported on similar systems~\cite{end10, var10}.
 \newline
 The intercalated noble metal atoms damp the G/Re moir\'e pattern as evidenced by low-energy electron diffraction measurements shown in Figure~1(h) and  also affect the core states of both graphene and Re. 
Previous photoemission studies of graphene on Re~\cite{min11} already reported that the C 1{\it s} peak (black full curve in Figure~\ref{arpes&cores}(a))  exhibits two main contributions centered at binding energies 285.05 and 284.45~eV. This splitting is a consequence of the strong buckling of the graphene film (about 1.6~\AA~according to DFT calculations~\cite{min11}), 
as the large corrugation defines differently interacting regions of the moir\'e  depending on the carbon positions relative to the substrate atoms. This leads to distinct C 1{\it s} components in the photoemission spectra, with the strongest interacting areas displaying contributions on the higher binding energy side of the spectra. On Rh(111), graphene displays two main C 1{\it s} peaks separated by 0.53~eV whereas on Ru(0001) the two main contributions are 0.60~eV apart~\cite{pre08}. The energy splitting of the C 1{\it s} peaks thus suggests that the corrugation of graphene on Re(0001) is comparable to that of graphene/Rh(111) and graphene/Ru(0001)~\cite{pre08}. Figure~\ref{arpes&cores}(a) also displays a residual carbide-like species at about 283~eV formed during the graphene growth. 
We estimate the amount of carbide to be smaller than 10 \% of the whole C at the surface and thus weakly affecting the electronic structure of the system.
\newline Intercalation of Ag (red dotted curve in Figure~\ref{arpes&cores}(a))  reduces the width of the whole C 1{\it s} structure by about 140~meV and shifts the center of the peak by $\sim$400~meV towards lower binding energies. Both effects may be ascribed to a reduced corrugation of the graphene film and a weaker C-substrate interaction induced by Ag. In addition, the coalescence of Ag adatoms on graphene and small dishomogeneity of the intercalated layer may also affect the energy and line-shape of the C 1{\it s} peak.
 \newline Figure~\ref{arpes&cores}(b) displays the photoemission spectrum of Re 4{\it f} states for the clean surface (yellow dashed curve) split by spin-orbit interaction into the doublet 4{\it f$_{7/2}$} and 4{\it f$_{5/2}$} at  40.22~eV and 42.64~eV binding energy respectively, in agreement with earlier studies~\cite{wag99}.  Since the 5{\it d} band of Re is half filled~\cite{cit83}, the  4{\it f} states display a negligible surface core level shift. The formation of graphene widens the photoemission peaks (black full curve) due to the C interactions with the Re surface atoms. 
 While the adsorption of Ag on top of graphene/Re does not affect either the line-shape nor the energetic position of the Re 4{\it f} states  (not shown),  the intercalation of Ag promotes  Ag-Re chemical bonds, resulting in a shift 
of the Re 4{\it f} peaks of the topmost layer by 0.6~eV with respect to the bulk components (red dotted curve).

 Additional effects induced by Ag intercalation are observed in
Figure~\ref{arpes&cores}(c) displaying the first derivative of a close-up of Figure~1(c). Four features  (labelled A, B, C and D) are identified, as well as a
band gap  in the graphene $\pi$ states (highlighted by the two arrows), similarly to graphene/Au/Ni~\cite{mar12, var08, shi13, pop11} and graphene/Ir(111)~\cite{bar12, kra11}.
{The character of these electronic states was analysed on the basis of DFT calculations performed for a single pseudomorphic Ag layer on the 
Re(0001) surface, which is consistent with the experimentally determined structure~\cite{par97}. 
In Figure~\ref{arpes&cores}(d), several dispersing Ag {\it d} bands with different symmetry located within the band gap of Re around the $\Gamma$ point are displayed.
At the $\Gamma$ point, the calculations predict two degenerate  {\it d$_{xz}$} and  {\it d$_{yz}$} states (E1) 
at  -3.0~eV, a  {\it s},{\it d$_{z^2}$} hybrid state at -3.8~eV, two degenerate  {\it d$_{xy}$} and  {\it d$_{x^{2}-y^{2}}$} states (E2)
at -4.3~eV and another   {\it s},{\it d$_{z^2}$} hybrid state at -6.0~eV below {\it E$_{F}$}. As one proceeds towards the ${\rm K}$ point the two-fold degeneracies are lifted
and the  {\it d$_{x^{2}-y^{2}}$} mixes into the  {\it s},{\it  d$_{z^2}$} hybrid opening a gap of $\approx$ 1~eV about midway between  $\Gamma$${\rm K}$. 
In comparison, the Ag {\it d} like single-particle DFT eigenvalues are located at significantly higher values than the experimental features.
Indeed the evaluation of the many-body effects on a G$_0$W$_0$ level [Fig.~\ref{arpes&cores}(e)] shows a shift of the bands to lower energies.
Yet it should be noted the many-body effects do not lead to a {\it constant} shift of all Ag {\it d} bands, 
but the shift of -1.4~eV for the  Ag E1 and
E2 bands is significantly more pronounced than
for the Ag  {\it s},{\it d$_{z^2}$} hybrid (-1.1~eV).
Recently, an energetic alignment of the silver bands with respect
to the substrate states has been proposed on the basis of a phase
accumulation model~\cite{plet08}.  
In the present work, the energetic position of the Ag {\it d} states on Re was evaluated on a true {\it ab-initio} level within the  GW framework.
We find that the quasiparticle effects on the band structure are in very good agreement with the effects predicted for bulk Ag \cite{mari}, indicating that in the present case the correlation effects are independent of the substrate. On the basis of these results, we may assign the features observed in the experimental spectra 
(Figure~\ref{arpes&cores}(c)) as follow}: A and C correspond to the Ag E1 ({\it d$_{xz}$},  {\it d$_{yz}$}) and Ag E2  ({\it d$_{xy}$},  {\it d$_{x^{2}-y^{2}}$}) states,
while the experimental bands B and D correspond to the to the two Ag {\it s},{\it d$_{z^2}$} bands. 

\begin{figure} [ht]
\includegraphics[width = \columnwidth]{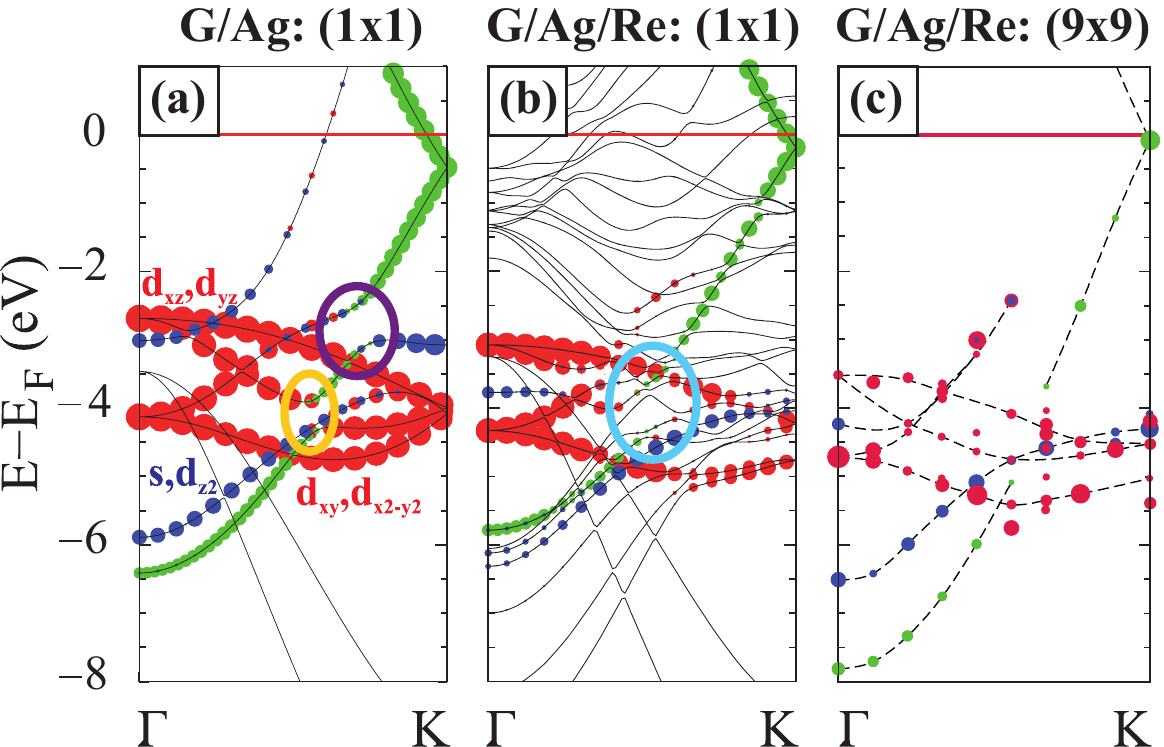}
\caption{(Color online) Band structure (vdW-DF, model structure) along the $\Gamma$${\rm K}$ direction (a) for graphene on a single ML Ag (G/Ag), (b) for graphene/Ag/Re (G/Ag/Re) in a (1$\times$1) model system and (c)
the unfolded band structure of a large (9$\times$9) model (bulk Re bands not shown). Bands localized in the Ag layer with symmetries {\it d$_{xz}$}, {\it d$_{yz}$}, {\it d$_{xy}$}, and  {\it d$_{x^{2}-y^{2}}$} are shown in red and with {\it s} and d$_{z^2}$ like character in blue. The localization in the respective layer is indicated by the dot-size. The $\pi$ band of graphene is displayed in green. Circles identify distinct electronic band gaps of graphene bands.
}
\label{DFT}
\end{figure}

To facilitate the discussion, the interaction of the graphene states with the substrate is analyzed in three steps, namely with an unsupported Ag layer, an epitaxic (1$\times$1) Ag/Re layer (which allows to seperate the contribution of the moir\'e lattice),
and for the full ($9 \times 9$) model.
Figure~\ref{DFT}(a) displays the band structure of a pseudomorphic graphene film adsorbed at 3.4~\AA~ on an {\it unsupported} single Ag layer (using the Re lattice parameter).  
These results indicate that the (almost) linear dispersion of graphene is preserved close to the Fermi level, but also show a significant hybridization between the graphene $\pi$ band and the 
Ag {\it d$_{xz}$} state at -4~eV (highlighted by the orange circle) and with the {\it d$_{x^{2}-y^{2}}$} and {\it d$_{z^2}$} states at about -3~eV (magenta circle), leading to the formation of {\it two} band gaps.
A comparison with the band structure of graphene/Ag/Re in Figure~\ref{DFT}(b) shows how this hybridization is modified by the Re substrate:  
As the Ag {\it d$_{xz}$} and  {\it d$_{z^2}$} states already hybridize with the Re states, the interaction with the Ag bands induces only 
a {\it single} band gap  (cyan circle) in the energy region in between 3-5~eV below E$_{F}$, consistent with the experimental observations. 
The origin of a related gap on G/Ir(111) has been recently discussed~\cite{bar12, kra11}, however in the present case, the observed band gap is mainly induced by the hybridization between the graphene  $\pi$ orbital and selected Ag {\it d} states.
It should be noted that the size of the graphene band gap is underestimated in the ($1 \times 1$) model structure due to the expanded graphene lattice: 
A comparison with the graphene band structure obtained for the more realistic ($9 \times 9$) model (Figure~\ref{DFT}c) shows 
a significant widening of the $\pi$ band in comparison to the smaller model, mostly visible at the bottom of the band:
At the {\rm K} point, both simulations predict a marginal {\it e}-doping with the Dirac point pinned at 0.12~eV and 0.18~eV below {\it E$_{F}$} in the ($9 \times 9$) and ($1 \times 1$) models, respectively.
However, at the $\Gamma$ point the lower edge of the graphene $\pi$ states is located at an energy of -5.9 eV  in the ($1 \times 1$) model,
while it is located nearly 2 eV lower in the more realistic ($9 \times 9$) cell. 
Furthermore, although the resolution in reciprocal space is limited by the computational costs, the almost vanishing localization of the graphene $\pi$ band 
when crossing the Ag $\it d$ states (Figure~3(c)) confirms a hybridization induced gap.
In conclusion, both models confirm the formation of a band gap due to  
 the hybridization of the graphene $\pi$ band  with the proper Ag {\it d} states, despite the rather large graphene-Ag/Re distance.

\section{Conclusion}

In summary, we have investigated the electronic structure of a graphene film adsorbed on bare and Ag monolayer covered Re(0001) surface
by means of angle-resolved photoemission experiments and DFT calculations. While the Ag layer results in a weaker  graphene-substrate interaction and 
restores the linear character of the $\pi$ band in proximity of {\it E$_{F}$}, 
we still find a significant hybridization with the Ag {\it d} bands at lower energies resulting in the formation of 
a band gap in the graphene $\pi$ band. 
The results clearly indicate that the ``weakly" interacting graphene sheet on the intercalated noble metal layer
is still electronically not completely decoupled. Hence the electronic structure of graphene adsorbed on a noble metal layer can still deviate significantly from 
the structure of an ideal, unsupported graphene sheet.

\subparagraph{Acknowledgements}

This work has been supported by the European Science Foundation (ESF) under the EUROCORES project EUROGRAPHENE/SpinGraph,
by the Austrian Science Fund (FWF) project I422-N16 and by the Italian Ministry of Education, University, and Research (FIRB Futuro in Ricerca, 
PLASMOGRAPH Project). The Vienna Scientific Cluster (VSC) is acknowledged for CPU time. Y.S. Kim and G. Kresse (Univ. Vienna) are gratefully acknowledged
for providing the routines for the unfolding of the electronic band structure.
 
$^{\ast}${e-mail:marco.papagno@fis.unical.it}

\end{document}